\documentstyle[preprint,aps,epsf]{revtex}

\newcommand{\nn}{\nonumber}
\newcommand{\LQCD}{\Lambda_{QCD}}
\newcommand{\asb}{\alpha_s^2\beta_0}

\newcommand{\al}{\alpha}
\newcommand{\bL}{\bar\Lambda}
\newcommand{\mB}{\overline{m}_B}
\newcommand{\basb}{\bar\alpha_s\beta_0}
\newcommand{\dd}{ {\rm d} }
\newcommand{\m}{\hat{\mu}}
\newcommand{\q}{ \hat{q}^2 }
\newcommand{\hq}{ \hat{q}^2 }
\newcommand{\hp}{\hat{P^0}}
\newcommand{\hz}{\hat{P}^2}
\newcommand{\hzz}{\hat{P}^4}
\newcommand{\GeV}{{\rm GeV}}
\newcommand{\an}{\bar{\nu_{\ell}}}

\begin{document}

\draft
 
{\tighten
\preprint{\vbox{\hbox{CALT-68-2075} }}
 
\title{ Order $\al_s^2\beta_0$ Correction to the Charged Lepton Spectrum 
in $b\to c\ell\an$ decays}
 
\author{Martin Gremm and Iain Stewart}

\address{California Institute of Technology, Pasadena, CA 91125, USA}

\maketitle

\begin{abstract}
We compute the $\asb$ part of the two-loop QCD corrections to the charged lepton
spectrum in $b\to c\ell \an$ decays and find them to be about 50\% of
the first order corrections at all lepton energies, except those close to the
end point.
Including these corrections we extract the central values $\bL=0.33\GeV$ and
$\lambda_1=-0.17\GeV^2$ for the HQET matrix elements and use them to 
determine the $\overline{\rm MS}$ $b$ and $c$ quark masses, and $|V_{cb}|$.

\end{abstract}

}

\newpage

\section{Introduction}
 In the last few years numerous theoretical and experimental
studies have focused on the electron spectrum in semileptonic inclusive $B\to
X_c\ell\an$ decays. 
The electron spectrum from free quark decays receives both
perturbative and nonperturbative corrections.
Knowledge of the shape of the spectrum can provide insights into
nonperturbative effects in $B$ meson decays, and thereby also
give some information on the weak mixing angle $|V_{cb}|$.
In the framework of  Heavy Quark Effective Theory (HQET) it is possible to show
that the quark level decay rate is the first term in a power series expansion
in the small parameter $\LQCD/m_b$ \cite{CGG}.
For infinitely heavy quarks the free quark model is an exact description of
heavy meson physics. At finite quark masses the first few terms in the heavy quark
expansion have to be taken into account. Expressions for these nonperturbative
corrections to the lepton spectrum are known to
order $(\LQCD/m_b)^3$ \cite{mark,mannel,bd,gk} and the ${\cal O}(\al_s)$
perturbative
corrections to the free quark decay were given in \cite{jk}.

The dominant remaining uncertainties are the two-loop corrections to the quark
level decay rate and the perturbative corrections to the coefficients of
the HQET matrix elements in the operator product expansion.
Here we examine the former.  While a full two-loop calculation of the electron
spectrum is a rather daunting task, it is possible to calculate the piece of the
two-loop correction that is proportional to $\beta_0=11-2/3 n_f$ with relative
ease by performing the one-loop QCD corrections with a massive gluon.  The
$\asb$ parts of the two-loop correction may then be obtained from a
dispersion integral over the gluon mass\cite{sv}. 
If there are no gluons in the tree level graph, the $\asb$ part of the
two-loop contribution is believed to dominate the full $\al_s^2$
result because $\beta_0$ is rather large. Several examples supporting this
belief are listed in \cite{asmark}.

A recent calculation \cite{asmark} of the $\al_s^2\beta_0$ correction to
the total inclusive rate for $B\to X_c\ell\an$ decays showed that the $\asb$
parts of the two-loop correction are approximately half as big as the
one-loop contribution, resulting in a rather low BLM scale \cite{BLM} of
$\mu_{BLM}=0.13m_b$.  For the electron spectrum we find that this part
of the second order
correction also amounts to about 50\% of the order $\al_s$ contribution,
at all electron energies except those close to the endpoint.
Close to the endpoint the corrections are roughly equal in magnitude.

The HQET matrix elements $\lambda_1$ and $\bL$ can be extracted from the
electron
spectrum in $B\to X_c\ell\an$ decays \cite{gklw}.  Even though this method of
obtaining HQET matrix elements was found to be rather insensitive to the first
order perturbative corrections, it is useful to extract $\lambda_1, \bL$
including the $\asb$ corrections. Then these matrix elements can be used to
relate the pole quark mass
to the $\overline{\rm MS}$ masses at order $\asb$. Similarly, one can include the
$\asb$ parts of the two-loop contribution in the theoretical prediction for the total
rate, which is needed for the determination of $|V_{cb}|$.  Since the $\asb$
corrections are rather large the resulting changes in the quark masses and
$|V_{cb}|$ are not negligible.

In Sect.~II and III we give analytic
expressions for the contributions from virtual and real gluon radiation. 
The last phase space integral in the virtual correction and
the last two integrals in the bremsstrahlung are done numerically. Readers not
interested in calculational details are advised to skip these sections.
In Sect.~IV we combine the results from the
previous two sections to obtain the $\asb$ corrections to the electron 
spectrum, and discuss
the implications for the extraction of $\bL,\lambda_1$, the $\overline{\rm MS}$
quark masses, and $|V_{cb}|$. In Appendix \ref{ap2} we give an interpolating
polynomial which reproduces the two-loop correction calculated here.

\section{Virtual Corrections}

The corrections from massive virtual gluons can be calculated in complete analogy
to the usual one-loop QCD corrections.  The ultraviolet divergence in the vertex
correction cancels when combined with the quark wave function renormalizations. 
There is no infrared divergence since we do the calculation with a massive gluon.
The virtual one-loop correction to the differential rate can be written as
\begin{eqnarray}
\frac{ \dd \Gamma^{(1)}_{virt}(\m) }{\dd y}&=&\al_s^{(V)}
\frac{|V_{cb}|^2G_F^2 m_b^5}{48 \pi^4}\int\dd \q
\Big[ 2(y-\q)(\q+1-r^2-y) (a_1+a_{wr}) -2 r \q a_2 \nn \\
&&+(\q(y-1)+y(1-r^2)-y^2)a_3\Big]  \label{v1loop}
\end{eqnarray}
where $\m=\mu/m_b$ is the rescaled gluon mass, and $y=2E_e/m_b$, $r=m_c/m_b$, and
$\q=q^2/m_b^2$ are the rescaled electron energy, charm mass, and momentum
transfer, respectively. The limits for the integration over $\q$ are
\begin{equation}
0\le\q\le\frac{y(1-y-r^2)}{1-y}, \qquad 0\le y\le 1-r^2.
\end{equation}
The functions $a_{wr}(\q)$ and $a_i(\q),\ i=1,2,3$ are the contributions from
the wave function renormalization and the vertex correction respectively.
They can be expressed in terms of 
the scalar two- and three-point functions $B_0$ and $C_0$ \cite{TV}, and the 
derivative $B^\prime_0=\partial B_0(a,b,c)/\partial a$.  Explicit expressions for 
these functions are given in Appendix \ref{ap1}.
Using the standard decomposition for the vector and tensor loop integrals \cite{TV}
we obtain
\begin{eqnarray}
a_1 &=& -2+4C_{00}+2(C_{11}+C_1+r^2C_{22}+r^2C_2)+2(1-\q+r^2)(C_{12}+C_0+C_1+C_2)\nn,\\
a_2 &=& 2r(C_1+C_2), \qquad
a_3 = -4(C_{11}+C_{12}+C_1)-4r^2( C_{12}+C_{22}+C_2),\\
a_{wr}&=&\frac{1}{2}\Bigg[ 2-B_0(1,\m^2,1)-B_0(r^2,\m^2,r^2)
+(1-\m^2)\Big(B_0(1,\m^2,1)-B_0(0,\m^2,1)\Big)\\
&&+\frac{(r^2-\m^2)}{ r^2} \Big(B_0(r^2,\m^2,r^2)-B_0(0,\m^2,r^2)\Big)
+2(2+\m^2)B^\prime_0(1,\m^2,1)\nn \\
&&+2(2r^2+\m^2)B^\prime_0(r^2,\m^2,r^2)\Bigg] \nn .
\end{eqnarray}
Defining $f_1=1+r^2-\q$ and $f_2=(f_1^2-4r^2)$ the coefficient functions take
the form
\begin{eqnarray}
C_{00}&=&\frac{1}{4f_2}\Bigg[ f_2
+\m^2(f_1-2)B_0(1,1,\m^2)+\m^2(f_1-2r^2)B_0(r^2,r^2,\m^2)\nn\\
&&+(f_2+2\q \m^2) B_0(\q,1,r^2)+2\m^2(f_2+\q \m^2)C_0(1,\q,r^2,\m^2,1,r^2)\Bigg],\\
C_{11} &=& \frac{r^2}{f_2}+\frac{(f_1-2r^2)(1-r^2)}{2\q f_2}B_0(0,1,r^2)
+\frac{f_1(\m^2-1)}{2f_2}B_0(0,1,\m^2)\\
&&+ \frac{3r^2\m^2(f_1-2r^2)}{f_2^2}B_0(r^2,r^2,\m^2)+
\frac{2\q \m^2(f_2+6\q r^2)-f_2(f_2+2\q r^2)}{2\q f_2^2}B_0(\q,1,r^2)\nn\\
&&+ \frac{\m^2 \Big( 6r^2(f_1-2)-f_2(f_1+2)\Big) } {2f_2^2}B_0(1,1,\m^2)\nn\\
&&+ \frac{2 \m^2 r^2 f_2+\m^4(f_2+6\q r^2) }{f_2^2}C_0(1,\q,r^2,\m^2,1,r^2)\nn,\\
C_{22}&=&\frac{1}{f_2}+\frac{(1-r^2)(2-f_1)}{2\q f_2}B_0(0,1,r^2)
+\frac{2\q \m^2(f_2+6\q)-f_2(f_2+2\q)}{2\q f_2^2}B_0(\q,1,r^2)\nn\\
&&+\frac{(\m^2-r^2)f_1}{2r^2f_2}B_0(0,r^2,\m^2)
+\frac{3\m^2(f_1-2)}{f_2^2}B_0(1,1,\m^2)\\
&&+\frac{\m^2\Big(6r^2(f_1-2r^2)-f_2(f_1+2r^2)\Big)}{2r^2f_2^2}B_0(r^2,r^2,\m^2)\nn\\
&&+\frac{\m^2\Big( 2f_2+\m^2(f_2+6\q)\Big)}{f_2^2}C_0(1,\q,r^2,\m^2,1,r^2) \nn ,\\
C_{12}&=&\frac{-f_1}{2f_2}+\frac{(1-r^2)(2r^2-f_1)}{2\q f_2}B_0(0,1,r^2)
+\frac{r^2-\m^2}{f_2}B_0(0,r^2,\m^2)\nn\\
&&+\frac{\m^2\Big(6(f_1-2r^2)-f_2\Big)}{2f_2^2}B_0(1,1,\m^2)
+\frac{\m^2\Big(6r^2(f_1-2)-f_2\Big)}{2f_2^2}B_0(r^2,r^2,\m^2)\\
&&+\frac{f_2(f_2+\q f_1)-2 \m^2\q(3\q f_1+f_2)}{2\q f_2^2}B_0(\q,1,r^2)\nn\\
&&+\frac{\m^2\Big(-f_1 f_2-\m^2(3\q f_1 +f_2) \Big)}{f_2^2}C_0(1,\q,r^2,\m^2,1,r^2)\nn,\\
C_1&=&\frac{1}{f_2}\Bigg[f_1  B_0(1,1,\m^2)+(2r^2-f_1)B_0(\q,1,r^2)
-2r^2B_0(r^2,r^2,\m^2)\nn \\
&&+\m^2(2r^2-f_1)C_0(1,\q,r^2,\m^2,1,r^2)\Bigg],\\
C_{2}&=&\frac{1}{f_2}\Bigg[ -2 B_0(1,1,\m^2)+(2-f_1)B_0(\q,1,r^2)+f_1B_0(r^2,r^2,\m^2)\nn \\
&&+\m^2(2-f_1)C_0(1,\q,r^2,\m^2,1,r^2)\Bigg].
\end{eqnarray}
The infinite parts of the regularized two-point functions can be shown to cancel in
eq. (\ref{v1loop}). In the limit $\m\to 0$ the vertex correction diverges
logarithmically. This divergence will be canceled by corresponding divergences in the
bremsstrahlung contributions discussed in the next section.

\section{Bremsstrahlung}

 The bremsstrahlung correction is found in the usual manner, by inserting a real
gluon on the $c$ and $b$ quark lines.   The calculation here is complicated by the
four-body phase space with two massive final states.  We follow the standard
procedure of decomposing the four-body phase space into a two- and a three-body 
phase space by introducing the four-momentum $P=p_c+p_g$. In the rest frame of the 
$b$ quark this decomposition reads
\begin{equation}
    dR_4 = dP^2\ dR_3(m_b; p_e, p_{\bar{\nu}}, P) dR_2(P; p_c, p_g).
\end{equation}
The ${\mathcal{O}}(\alpha_s)$ bremsstrahlung correction to the
differential rate is given in terms of dimensionless variables ($\hp=P^0/m_b$, 
$\hz=P^2/m_b^2$) by
\begin{eqnarray}
 {d \Gamma_{brems}^{(1)}(\m) \over \dd y } &=& \al_s^{(V)}{G_F^2\, |V_{cb}|^2\, m_b^5 \over 192\ \pi^4 }\
  \int d\hz\ d\hp\  (\hp^2-\hz)^{-5/2} \biggl[ 2 b_1 (1-2\hp+\hz) \nonumber \\
   & & \qquad + b_2 (2-2\hp-y) y + b_3 (1-y-\hz) (2\hp+y-\hz-1) \nonumber \\
  & & \qquad  + b_4 (1-y-\hz) y +  b_5 (2\hp +y-2) (1-2\hp-y+\hz) \biggl] . 
   \label{r1loop}
\end{eqnarray}
For convenience the above rate has been written in terms of the 
coefficients $b_i$
\begin{eqnarray}
  b_1 &=& (\hp^2-\hz) \left( \hz(c_2-c_1)+\hp^2 c_1+c_3-\hp(c_4+c_5) \right),\\
  b_2 &=& (\hp^2-\hz)\hz c_1 +3\hzz c_2+(\hz+2\hp^2) c_3 -3\hp\hz (c_4+c_5) , \\
  b_3 &=& (\hp^2-\hz) c_1 + (\hz+2\hp) c_2+3 c_3 - 3 \hp (c_4+c_5) ,  \\
  b_4 &=& -\hp(\hp^2-\hz)c_1-3\hp\hz c_2-3\hp c_3+(\hp^2+2\hz) c_4+3\hp^2 c_5,\\
  b_5 &=& -\hp(\hp^2-\hz) c_1-3\hp\hz c_2-3\hp c_3 + 3\hp^2c_4 + (\hp^2+2\hz) c_5 , 
\end{eqnarray}
which are linear combinations of 
\begin{eqnarray}
  c_1 &=& {4 (v_+^2 - v_-^2) \over h} 
            + { 2 [ (h+\m^2-2\hp)^2+(\m^2-2\hp)^2+2 \m^2 (1+r^2) ] \over 
             h} \ln( {2 v_+ -\m^2 \over 2 v_- - \m^2}) \nonumber \\
  \quad &+& {4 (2+\m^2)(h-2\hp)(v_+-v_-) \over (2 v_+ -\m^2) (2 v_- -\m^2) }
        - {8 [(z+\m^2)(\hp-h)+z\hp] (v_+-v_-) \over h^2 } , \\
  c_2 &=& {2 (v_+^2 - v_-^2) (2-h) \over h} 
           - { [h \m^2(3\m^2-4\hp)+4\m^2(2\hp-1)-16\hp^2] \over h} 
                \ln( {2 v_+ -\m^2 \over 2 v_- - \m^2})  \nonumber \\
   \quad &+& {2 (\m^4-4)(2\hp-\m^2)(v_+-v_-) \over (2 v_+ -\m^2) (2 v_- -\m^2) }
           - { 4 [h^2 (\m^2-\hp)+2\hp (\m^2+2\hz)] (v_+-v_-) \over h^2 } , \\
  c_3 &=& {[(\hz+r^2) (h^2+2(\m^2-2\hp)(h-2\hp)) -h\m^4 +4 r^2 \hz \m^2] \over h} 
                \ln( {2 v_+ -\m^2 \over 2 v_- - \m^2}) \nonumber \\
   \quad &+& {2 (2+\m^2) (\m^2-r^2-\hz)(2\hp-\m^2-h) (v_+-v_-)
              \over (2 v_+ -\m^2) (2 v_- -\m^2) } \nonumber \\
   \quad &-& {4 [(\hz+r^2)(h\hp-h z+\m^2\hp)+4r^2\hz\hp] (v_+-v_-) \over h^2 } , \\
  c_4 &=& {4 (\hp-\hz) (v_+^2 - v_-^2) \over h} + {2 (2+\m^2)(2\hp-\m^2)(\m^2-2\hp+h)
            (v_+-v_-)\over(2 v_+ -\m^2)(2 v_- -\m^2) } \nonumber \\
   \quad &-& {2 [4\hp^2(2\hz+\m^2)+(\m^2-r^2) h (2\hz+h-4\hp)+h\hz(\hz+r^2-8\hp)] 
         (v_+-v_-)  \over h^2 } \nonumber \\
   \quad &-& {2 \over h}  [2(\m^2-\hp)(h \m^2-2\hp h+4\hp^2)+h^2(1-\hp+\m^2)-
           2\m^2(r^2+r^2\hp+\hp) \nonumber \\ 
         &+& \m^4(1+r^2-2\hp)]  \ln( {2 v_+ -\m^2 \over 2 v_- - \m^2}) , \\
   c_5 &=& {2 (\m^4-4) (\hz+r^2-\m^2)(v_+-v_-)\over(2 v_+ -\m^2)(2 v_- -\m^2)} \nonumber \\
         &-& {2 [\m^2h^2+(\hz+r^2)(2\m^2+2h-h^2)+8r^2\hz](v_+-v_-)\over h^2} \nonumber \\
         &-& {2 [\m^4h-(\hz+r^2)(\m^2h+4\hp)+2\hz \m^2] \over h} 
                \ln( {2 v_+ -\m^2 \over 2 v_- - \m^2}) .
\end{eqnarray}
In the expressions for the $c_i$ we have put $h=\hz-r^2$ and 
\begin{equation}
   v_\pm = { (\hz+\m^2-r^2)\hp \pm \sqrt{\hp^2-\hz} \sqrt{(\hz+\m^2-r^2)^2-4 \m^2\hz} 
    \over 2\hz }.
\end{equation}
The integrals in eq. (\ref{r1loop}) are done numerically between the kinematic
limits
\begin{eqnarray}
  {(1-y)^2+\hz \over 2 (1-y)} \le &\hp& \le {1+\hz \over 2} ,\\
(\m + r)^2 \le &\hz& \le 1-y.
\end{eqnarray}
To improve the numerical stability for small $\m^2$ we found it useful to do
 the $\hp$ integral with the variable $\ln(\hp-r^2)$.  The remaining limits for 
this four-body decay are
\begin{equation}
      0   \le   \m   \le  \sqrt{1-y} - r, \qquad  0 \le y \le 1-r^2.
\end{equation}

\section{The $\asb$ correction}

Combining the corrections from virtual and real gluon radiation,
eqs. (\ref{v1loop},\ref{r1loop}), we obtain 

\begin{equation}\label{glue}
\frac{\dd\Gamma^{(1)}(\m)}{\dd y} = \frac{\dd\Gamma^{(1)}_{virt}(\m)}{\dd y}
+\frac{\dd\Gamma^{(1)}_{brems}(\m)}{\dd y}\Theta(\sqrt{1-y}-r-\m).
\end{equation}
In the $\m\to 0$ limit eq. (\ref{glue}) yields the one-loop correction to the
electron spectrum.
We have checked that our expression reproduces the result in \cite{jk} in this
limit.  The $\asb$ part of the two-loop correction is related to the one-loop
expression calculated with a massive gluon by \cite{sv}
\begin{equation} 
\frac{\dd\Gamma^{(2)}}{\dd y}=-\frac{{\al_s^{(V)}}\beta_0}{4\pi}
\int\limits_{0}^{\infty}\frac{\dd \m^2}{\m^2}
\left( \frac{\dd \Gamma^{(1)}(\m)}{\dd y}-
\frac{1}{1+\m^2}\frac{\dd \Gamma^{(1)}(0)}{\dd y}\right).
\end{equation}
Note that $\al_s^{(V)}$, defined in the V-scheme of BLM \cite{BLM}, is related
to the more familiar $\bar\al_s$ defined in the $\overline{\rm MS}$ scheme by
\begin{equation}
\al_s^{(V)}=\bar\al_s+\frac{5}{3}\frac{\bar\al_s^2}{4\pi}\beta_0+\cdots.
\end{equation}
$\al_s$ is evaluated at $m_b$ unless stated otherwise.
In the $\overline{\rm MS}$ scheme the $\bar\asb$ part of the two-loop
correction reads
\begin{equation}
\frac{\dd\Gamma^{(2)}}{\dd y}=\frac{5}{3}\frac{\basb}{4\pi}
\frac{\dd \Gamma^{(1)}(0)}{\dd y}
-\frac{\basb}{4\pi}\int\limits_{0}^{\infty}\frac{\dd \m^2}{\m^2}
\left( \frac{\dd \Gamma^{(1)}(\m)}{\dd y}-
\frac{1}{1+\m^2}\frac{\dd \Gamma^{(1)}(0)}{\dd y}\right).
\end{equation}
The dispersion integral has to be done with some care.
We found that using $\ln(\m^2)$ instead of $\m^2$ as the
integration variable simplifies the numerical evaluation considerably.

In Fig. 1 we plot the $\bar\asb$ part of the two-loop correction\footnote{
In Appendix \ref{ap2} we give a polynomial fit to our results that
reproduces the numerical results to better than 1\% for mass ratios in the range
$0.29\le r\le 0.37$.}
and for comparison the one-loop correction to the electron spectrum,
using $r=0.29$, $\bar\al_s=0.2$, $n_f=3$, and dividing by
$\Gamma_0=G_F^2|V_{cb}|^2m_b^5/192\pi^3$.
Except for electron energies close to the endpoint, the $\asb$ corrections are
about half as big as the first order corrections. The perturbation series
appears to be controlled but the higher order corrections clearly are not
negligible.
Integrating over the electron energy we reproduce the result for the 
correction to the total rate given in Ref. \cite{asmark}.

In \cite{gklw} the HQET matrix elements $\bL,\lambda_1$ were extracted from
the lepton spectrum using the experimentally accessible observables
\begin{equation}
R_1=\frac{\int_{1.5\GeV} \dd E_\ell \, E_\ell \dd \Gamma/\dd E_\ell}
{\int_{1.5\GeV} \dd E_\ell \, \dd \Gamma/\dd E_\ell}, \qquad
R_2=\frac{\int_{1.7\GeV} \dd E_\ell \, \dd \Gamma/\dd E_\ell}
{\int_{1.5\GeV} \dd E_\ell \, \dd \Gamma/\dd E_\ell}.
\end{equation}
It is straightforward to calculate the $\bar\asb$ corrections to these
quantities. In the spirit of HQET, we use the spin averaged meson masses
$\overline{m}_B=5.314\GeV$, and $\overline{m}_D=1.975\GeV$
instead of quark masses. Keeping only two-loop corrections that are proportional
to $\beta_0$, and neglecting terms of order $\bar\asb\LQCD/\mB$ we find
\begin{eqnarray}\label{r1r2}
R_1&=&1.8059-0.035\frac{\bar\al_s}{\pi}-0.082\frac{\bar\asb}{\pi^2}+\cdots,\\
R_2&=&0.6581-0.039\frac{\bar\al_s}{\pi}-0.098\frac{\bar\asb}{\pi^2}+\cdots\nn,
\end{eqnarray}
where the ellipsis denote the other contributions including 
nonperturbative corrections discussed in \cite{gk,gklw}.
The BLM scales for these
quantities are $\mu_{BLM}(R_1) = 0.01 \mB$, and $\mu_{BLM}(R_2) = 0.007\mB,$
reflecting the fact that the second order corrections are larger than the first
order. This is a result of the almost complete cancellation of the first order
perturbative corrections from the denominators and numerators in $R_{1,2}$.
In eq.(\ref{r1r2}) the BLM scales for the numerators and denominators are
separately comparable to the BLM scale for the total rate
$\mu_{BLM}\approx 0.1\mB$. Therefore the very low BLM scales of $R_{1,2}$
do not necessarily indicate badly behaved perturbative series.

In order to demonstrate the impact of the $\bar\asb$ corrections on the
extraction of $\bL,\lambda_1$, we repeat the analysis of \cite{gklw} neglecting 
nonperturbative corrections of order $(\LQCD/m_b)^3$ which may be
substantial \cite{gk}. Because of the higher order nonperturbative 
corrections, large theoretical uncertainties have to be assigned to the extracted
values of $\bL,\lambda_1$.  We find that the central values are moved from
$\bL=0.39\pm0.11\GeV,\,\lambda_1=-0.19\pm 0.10 \GeV^2$ to
$\bL=0.33\GeV,\,\lambda_1=-0.17$.
The shift in the values of the HQET matrix elements lies well within the
$1\sigma$ statistical error of the previously extracted values.
However, using the values of the HQET matrix elements extracted at a given order
in $\al_s$ to predict physical observables at the same order in $\al_s$,
guarantees that the renormalon ambiguity in $\bL$ and $\lambda_1$ will
cancel\cite{renorm,renorm2} if the
expansion is continued to sufficiently high orders in $\al_s$. Thus
including the $\bar\asb$ parts in the determination of $\bL,\lambda_1$ 
allows one to calculate the $\overline{\rm MS}$ quark masses consistently
at order $\bar\asb$.
To second order in $\LQCD/m_q$ and to order $\bar\asb$ we have 
\begin{equation}
\overline{m}_q(m_q)=\left(\overline{m}_{Meson}-
\bL+\frac{\lambda_1}{2m_q}+\cdots\right)
\left(1-\frac{4\bar\al_s(m_q)}{3\pi} -1.56\frac{\bar\al_s^2(m_q)\beta_0}{\pi^2}+\cdots\right),
\end{equation}
where $m_q$ is the $b$ or $c$ quark pole mass and $\overline{m}_{Meson}$ is the
corresponding spin averaged meson mass. With $\bar\al_s(m_b)=0.22,\,\bar\al_s(m_c)=0.39$ this yields
$\overline{m}_b(m_b)=4.16\GeV, \overline{m}_c(m_c)=0.99\GeV$ for the 
 $\overline{\rm MS}$ quark masses, albeit
with large theoretical uncertainties due to the effect of the higher order
nonperturbative corrections on the extraction of $\bL,\lambda_1$\cite{gk}. 
The value of $\overline{m}_b(m_b)$ is in good agreement with lattice
calculations $\overline{m}_b(m_b)=4.17\pm 0.06\GeV$ and $\overline{m}_b(m_b)=
4.0\pm 0.01\GeV$\cite{lat}.
The weak mixing angle $|V_{cb}|$ can be determined by comparing the theoretical
prediction for the total rate with experimental measurements.
Including all corrections discussed in \cite{gklw} we find at order $\asb$ 
\begin{equation}
|V_{cb}|=0.043\left(\frac{ Br(B\to X_c\ell\an)}{0.105} \frac{ 1.55{\rm ps}}{\tau_B}\right)^{1/2}.
\end{equation}

\section{Conclusions}
We have calculated the ${\cal O}(\asb)$ corrections to the electron
spectrum in $b\to c\ell\an$ decays which turn out to be rather large,
about 50\% of the one-loop corrections. These corrections can be included in the
extraction of the HQET matrix elements $\bL,\lambda_1$. We obtain
$\bL=0.33\GeV$ and $\lambda_1=-0.17\GeV^2$, both somewhat lower than the
values extracted at ${\cal O}(\al_s)$. Using these values and including 
${\cal O}(\asb)$ corrections we obtain $\overline{m}_b(m_b)=4.16\GeV,\,
\overline{m}_c(m_c)=0.99\GeV$ for the $\overline{\rm MS}$ quark masses and
$|V_{cb}|=0.043( Br(B\to X_c\ell\an)/0.105\times 1.55{\rm ps}/\tau_B)^{1/2}$.
These results have large theoretical uncertainties due to the effect of
nonperturbative corrections of order $(\LQCD/m_b)^3$ on the extraction of
$\bL,\lambda_1$.

\acknowledgments
We would like to thank Anton Kapustin, Zoltan Ligeti and Mark Wise for helpful
discussions. This work was supported in part by the Department of Energy
under grant DE-FG03-92-ER 40701.

\appendix
\section{Scalar two- and three-point Functions}
\label{ap1}
 Here we list expressions for the scalar functions $B_0$ and $C_0$ \cite{TV}
needed for the calculation in Section II.  With the cut for
logarithms along the negative real axis we have
\begin{eqnarray}
   B_0(a,b,c) &=& {2 \over \epsilon} - \ln({\mu^2 \over 4\pi\Lambda^2 e^{-\gamma}}) 
     - \int_{0}^{1} dx \ln( {a x^2 - x (a+b-c)+b \over \mu^2} )  \\
   &=& {2 \over \epsilon} - \ln({\mu^2 \over 4\pi\Lambda^2 e^{-\gamma}})
     + 2 - \ln( {c \over \mu^2} )
     + x_+ \ln( {x_+-1\over x_+}) + x_-\ln({x_--1\over x_-}),
\end{eqnarray}
where 
\begin{equation}
     x_{\pm} = { a+b-c \pm \sqrt{ (a+b-c)^2 -4ab } \over 2 a}.
\end{equation}
\begin{eqnarray}
  C_0(1,\hq,r^2&,&\m^2,1,r^2) = {1 \over 1+r^2-\hq-2\alpha r^2} \times \nonumber \\
& & \biggl\{ 
     Li_2 \left({z_1 \over z_1-z_4}\right) - Li_2 \left({z_1-1 \over z_1-z_4}\right)
    +Li_2 \left({z_1 \over z_1-z_5}\right) - Li_2 \left({z_1-1 \over z_1-z_5}\right) 
      \nonumber \\
 & &-Li_2 \left({z_2 \over z_2-z_6}\right) + Li_2 \left({z_2-1 \over z_2-z_6}\right)
    -Li_2 \left({z_2 \over z_2-z_7}\right) + Li_2 \left({z_2-1 \over z_2-z_7}\right)
      \nonumber \\
 & &+Li_2 \left({z_3 \over z_3-z_8}\right) - Li_2 \left({z_3-1 \over z_3-z_8}\right)
    +Li_2 \left({z_3 \over z_3-z_9}\right) - Li_2 \left({z_3-1 \over z_3-z_9}\right)
   \biggl\} ,
\end{eqnarray}
where
\begin{eqnarray}
   \alpha&=&{(\hq-1-r^2) \pm \sqrt{(\hq-1-r^2)^2-4r^2} \over -2r^2},\ \  
   z_0 = { \m^2-2- \alpha(\hq-1-r^2+\m^2) \over 1+r^2-\hq-2r^2\alpha},\\
   z_1 &=& z_0+\alpha,\quad z_2 = {z_0 \over (1-\alpha)},\quad
   z_3 = - {z_0 \over \alpha}, \quad
   z_{4,5} = {\m^2 \pm \sqrt{\m^4 -4 \m^2 r^2} \over 2 r^2},\\
   z_{6,7}&=&{\hq+1-r^2 \pm \sqrt{(\hq +1-r^2)^2 - 4\hq +i\epsilon} \over 2\hq} , \ \ 
   z_{8,9} = {2-\m^2 \pm \sqrt{(2-\m^2)^2-4} \over 2}.
\end{eqnarray}
The dilogarithms here are defined as
\begin{equation}
   Li_2(z) = - \int_{0}^{1} dt {\ln(1-z t) \over t}.
\end{equation}

\section{Interpolating Function}
\label{ap2}

  In this appendix we give an interpolation scheme that makes it easy to reproduce
our $\bar\asb$ corrections to the lepton spectrum for
$b\to c\ell\an$ decays.  The interpolation is done as a function of $y$ and $r$ with
all other parameters left explicit.
\begin{equation}
   \frac{\dd\Gamma}{\dd y}=\Gamma_0 \left( f^{(0)}(y,r) + \frac{\bar{\al}_s}{\pi}
        f^{(1)}(y,r) + \frac{\bar{\asb}}{\pi^2} f^{(2)}(y,r) \right).
\end{equation}
The well known tree level result is given by the exact equation
\begin{equation}
  f^{(0)}(y,r) = \frac{2y^2(y+r^2-1)^2(-3-3r^2+5y+r^2y-2y^2)}{(y-1)^3},
\end{equation}
while the first order function, $f^{(1)}$, can be obtained from Ref. 
\cite{jk}.  The interpolating function $f^{(2)}$ is found by 
fitting Chebyshev polynomials, $T_n$, to the energy dependence, and quadratic 
polynomials to the mass ratio $r$.  To improve the accuracy of the fit
near the endpoint of the lepton spectrum the expansion is given in terms of 
the variable 
\begin{equation}
  y' = \frac{\ln((1-y)^2/r^2)}{\ln(r^2)}.
\end{equation}
Our fit is accurate to better than 1\% for the region
\begin{equation}
  0.09  \le y \le 0.99 (1-r^2)  \qquad  0.29 \le r \le 0.37  .
\end{equation}
The second order function is given by 
\begin{eqnarray}
  f^{(2)}(y,r) = \sum_{n=1}^{12} 
       \left(\sum_{m=0}^{2} A_{n,m}\ r^m\right) \ T_{n-1}(y') ,
\end{eqnarray}
where $T_n(y') = \cos( n \arccos y')$.
The 36 coefficients $A_{i,j}$ are given in Table 1.  Both 
the coefficients as well as Fortran code evaluating $\dd\Gamma^{(2)}/\dd y$
and $\dd\Gamma^{(1)}/\dd y$ from the expressions in sections II,III, and IV 
are available from the authors by request (iain@theory.caltech.edu).

\begin{table}[t]
\caption{Coefficients $A_{n,m}$ for the interpolation function $f^{(2)}(y,r)$
given in Appendix \ref{ap2}. }
\rule{0in}{2ex}
\begin{center}
\begin{tabular}{rrrr}  
   $A_{n,m}$  & $m=0$ & $1$ & $2$  \\ \hline
  $n=1$  &  $-4.9$     &   $19.3$     &  $-20.4$  \\ 
  $2$    &  $-2.72$    &   $12.4$     &  $-14.7$  \\
  $3$    &   $\ 3.55$    &  $-12.4$     &   $11.5$  \\
  $4$    &   $\ 2.21$    &  $-10.1$     &   $11.9$  \\ 
  $5$    &   $\ 1.97$    &   $-9.59$    &   $11.9$  \\
  $6$    &   $\ 1.11$    &   $-5.08$    &    $6.06$  \\
  $7$    &  $-0.159$   &    $0.448$   &   $-0.274$ \\ 
  $8$    &  $-0.336$   &    $1.44$    &   $-1.62$  \\
  $9$    &  $-0.319$   &    $1.47$    &   $-1.76$  \\ 
  $10$   &  $-0.174$   &    $0.818$   &   $-0.994$  \\ 
  $11$   &  $-0.0881$  &    $0.417$   &   $-0.509$  \\ 
  $12$   &  $-0.0507$  &    $0.247$   &   $-0.309$  \\
\end{tabular}
\end{center}
\label{table_bn1}
\end{table}

\newpage
\begin{figure}
\begin{center}
\centerline{\epsfxsize=10.cm  \epsfbox{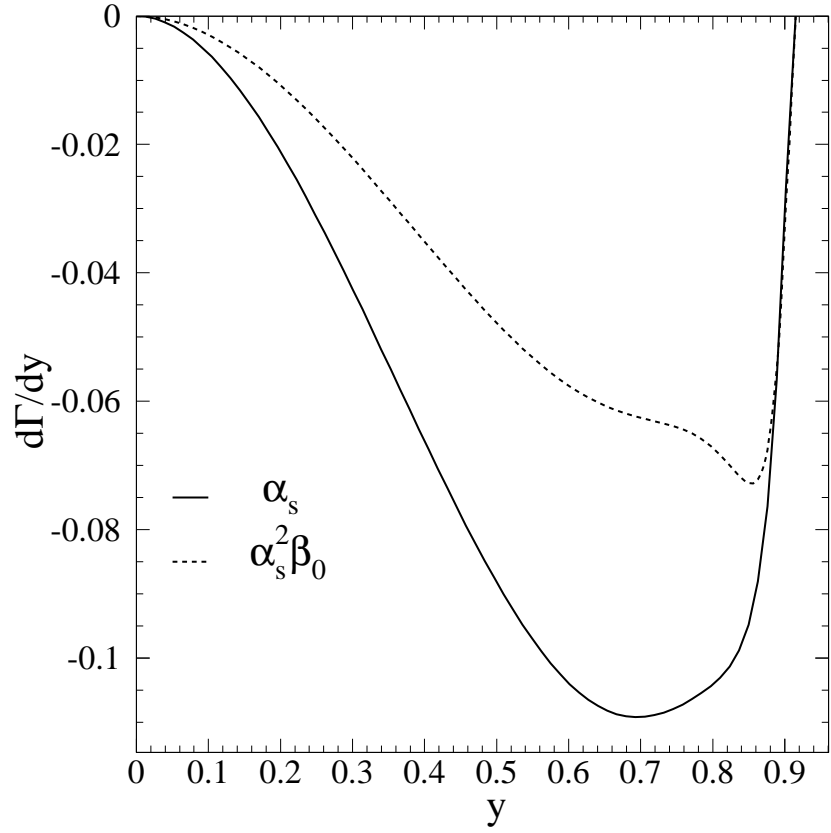}}
\caption{Perturbative QCD corrections to the lepton spectrum.  The differential rates
  are given in units of $\Gamma_0$, with $r=0.29$, $\bar{\al}_s=0.2$, and $n_f=3$.}
\label{fig_1}
\end{center}
\end{figure}

\end{document}